# Balancing rotating structures using slow-speed data via optimized parametric excitation and nonlinear feedback

A. Dolev[1]*, S. Tresser[1] and I. Bucher[1]*
[1] Dynamics Laboratory, Mechanical Engineering, Technion
Technion City, Haifa 3200003, Israel
*e-mail: **amitdtechnion@gmail.com, bucher@technion.com**

## Abstract
The paper presents an improved mass balancing procedure for fast rotating machinery, while it is being rotated at speeds considerably slower than the "critical speeds", where dangerously high vibration amplitudes may arise. By utilizing tuned dual frequency parametric excitation along with optimized nonlinear feedback terms, the slow imbalance forces are projected onto a chosen mode of vibration. This allows to identify the imbalance projection on that specific mode, and to cancel these forces by adding or reducing mass. The scheme benefits from two kinds of parametric excitation yielding combination and principal parametric resonances. The former is used to project the imbalance forces onto a selected vibration mode, and the latter significantly amplifies the response. By tuning the parametric excitation and the nonlinear terms in an optimal manner, a pseudo-linear behavior is formed. This behavior enables to increase both amplification and sensitivity to the imbalance forces without having to compromise between the two.

## 1    Introduction

Mechanical vibrations in rotating machinery is an undesired phenomenon, and its main excitation source is rotor imbalance. Because the local center of mass at each cross-section along the rotor does not coincide with the geometric center, a rotating distributed force arises at the rotation frequency. This force excites all the vibration modes of the structure, and each mode's participation in the response is proportional to the imbalance projection on it [1,2]. To minimize the imbalance, a mass balancing procedure (MBP) is often carried out, where small correction masses are added or removed. Usually, the MBP is performed at the machinery full operational speed range.

Rotors exhibiting bending oscillations at their operational speed range are referred to as "flexible", and are common in high speed machinery. This kind of machinery is customarily designed to operate close to a specific critical speed where a single "flexible mode" dominates the response. Implementation of MBP for such machinery is impossible in many cases due to safety reasons or harsh conditions, e.g. high temperatures or inaccessibility, which prevent the ability to take measurements or even to run the system at such speeds. The inability to perform flexible MBP may lead to conservative over-design, adding damping elements which add weight and unacceptable complexity [3,4], and performing MBP using commercial balancing machines [5], where the rotor is not balanced in its full operational speed range. The latter assumes a rigid rotor type of behavior, hence the projection of the imbalance on flexible modes is not seen by sensors. At the related critical speeds which are not balanced, large vibrations are exhibited during operation [6,7].

In all MBP approaches designed for flexible rotors (e.g., "Influence Coefficient Method", "Modal Balancing" and "Unified Balancing Approach" [6–14]), the machinery is span in the vicinity of critical speeds within its full operational speed range and the vibration levels are measured [6,8–10]. Then, these data are post-processed to calculate the correction masses. Since these methods are based on vibration readings, a sufficiently good signal to noise ratio is required. The latter is achieved when the vibration levels are sufficiently high, therefore, it is required to rotate the machinery close to its critical speeds [6]. Furthermore, when this occurs, a single vibration mode dominates the response, allowing separating its contribution to the measured response. This procedure has to be repeated for all modes of interest, therefore at several critical speeds. The latter cannot be carried out unless all critical speed in the range are balanced.

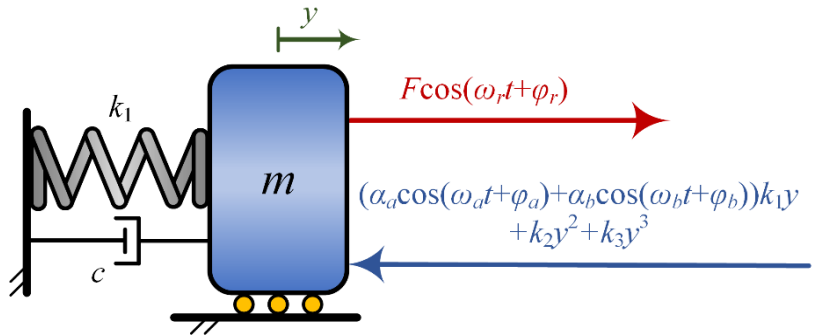

Figure 1: A linear SDOF oscillator subjected to a controlled force.

Recently, a different MBP approach was reported [15,16], based on [17–20], which enables to detect the imbalance projection on high frequency modes while rotating at speeds much lower than the critical speeds. The reported scheme allows to overcome the aforementioned limitations by utilizing dual-frequency parametric excitation (also known as pumping) along with a cubic nonlinear feedback term which limits the vibration levels and leads to a steady oscillation amplitude. The latter approach was implemented in [15,16], and it was shown that the imbalance can be found with high accuracy both numerically and experimentally. It was found in [16] that the suggested MPB is robust, however, when very weak imbalance forces are considered, high precision in the amplifier's parameters estimation is needed. Moreover, lower sensitivity levels were obtained with respect to cases with larger imbalance. This fact leads to a compromise, when tuning the parameters, between the amplification and sensitivity levels.

An improved scheme for which optimal parameters tuning is given in a closed form was recently introduced [21]. The scheme was developed for a nonlinear single-degree-of-freedom (SDOF) parametric amplifier, which behaves in a pseudo-linear manner. This unique dynamic behavior is achievable by a carful tuning of the cubic and quadratic nonlinear terms [21–23]. Furthermore, it eliminates the need to make a comprise between amplification and sensitivity levels, and permits to lower the precision requirement in the amplifier parameter estimation. In what follows, the scheme of [21] is extended to accommodate a multi-degree-of-freedom (MDOF) rotating machinery dynamic behavior. The ability to detect the imbalance force projection on any desired vibration mode, while rotating much slower than the critical speeds, without the aforementioned compromise is demonstrated via a test case.

The paper begins with a brief mathematical summary of the pseudo-linear parametric amplifier scheme, and the stages required to tune the amplifier parameters. Then, the improved MBP scheme is introduced, followed by a concise mathematical description of the scheme extension to a rotating MDOF system. In the following sections, a numerical verification of the suggested optimal scheme is carried out, and the performance is compared to the previously reported scheme.

## 2    Approximate analytical solution of a pseudo-linear parametric amplifier

A linear SDOF oscillator which is subjected to the aforementioned controlled external force is analyzed in this section. The discussion below is related to a oscillator and the nature of forces acting on it as illustrated in Figure 1. The model considered here consists of an ordinary differential equation having lumped parameters and is characterized by a point mass $m$, linear dashpot $c$ and a linear stiffness $k_1$. The applied force consists of three components: (1) a dual-frequency parametric excitation term with frequencies $\omega_a$, $\omega_b$, appropriate phase shifts $\varphi_a$, $\varphi_b$ and corresponding magnitudes $\alpha_a$, $\alpha_b$, (2) nonlinear stiffness feedback terms – quadratic ($k_2$) and cubic ($k_3$), and (3) an external force at a known frequency, $\omega_r$, and unknown phase shift $\varphi_r$ and amplitude $F$. The third term, the harmonic force (at frequency $\omega_r$), is the input signal to be amplified.

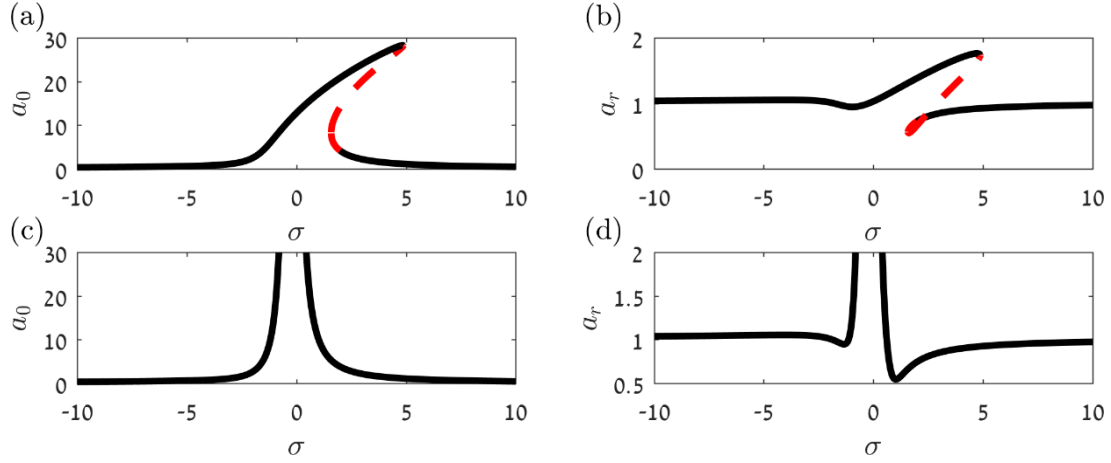

Figure 2: (a) The amplitude of the harmonic term near the natural frequency and (b) at the external force frequency vs. the detuning parameter when $\kappa_e = 5$. (c) The amplitude of the harmonic term near the natural frequency and (d) at the external force frequency vs. the detuning parameter, when $\kappa_e = 0$. Continuous lines represent stable solutions while dashed lines represent unstable solutions.

The governing equation of motion is:

$$m\ddot{y} + c\dot{y} + \left(\left(1 + \alpha_a \cos(\omega_a t + \varphi_a) + \alpha_b \cos(\omega_b t + \varphi_b)\right)k_1 + k_2 y + k_3 y^2\right) y = F \cos(\omega_r t + \varphi_r). \quad (1)$$

The analytical solution is approximated by utilizing the multiple scales perturbation method [24], therefore the equation of motion is transformed to the following:

$$x'' + x = P\cos(\Omega_r \tau + \varphi_r) - \varepsilon\left(2\zeta x' + \left(\gamma_a \cos(\Omega_a \tau + \varphi_a) + \gamma_b \cos(\Omega_b \tau + \varphi_b)\right)x + \kappa_2 x^2\right) - \varepsilon^2 \kappa_3 x^3$$

$$x = y/\varepsilon^2 \quad \tau = \omega_n t, \quad \omega_n = \sqrt{k/m}, \quad \Omega_\bullet = \frac{\omega_\bullet}{\omega_n}, \quad \zeta = \frac{c}{2\varepsilon m \omega_n}, \quad (2)$$

$$\gamma_\bullet = \alpha_\bullet / \varepsilon, \quad \kappa_2 = \frac{\varepsilon k_2}{m\omega_n^2}, \quad \kappa_3 = \frac{\varepsilon^2 k_3}{m\omega_n^2}, \quad P = \frac{F}{m\varepsilon^2 \omega_n^2}.$$

Whereas, $\partial \bullet / \partial \tau \triangleq \bullet'$, and it is assumed that the amplifier is lightly damped $\varepsilon \sim O(c/2\sqrt{km}) \ll 1$. Considering a second order expansion in the form

$$x(\varepsilon, \tau) = x_0(\tau_0, \tau_1, \tau_2) + \varepsilon x_1(\tau_0, \tau_1, \tau_2) + \varepsilon^2 x_2(\tau_0, \tau_1, \tau_2), \quad \tau_i = \varepsilon^i \tau, \quad (3)$$

and setting the pumping frequencies as:

$$\Omega_a = 2(1 + \varepsilon\sigma), \quad \Omega_b + \Omega_r = 1 + \varepsilon\sigma, \quad (4)$$

whereas $\sigma$ is a detuning parameter. The selected frequencies in Eq.(4) mean pumping the system at a frequency ($\Omega_a$) close to twice the natural frequency yielding a principal parametric resonance [25], and a frequency combination ($\Omega_b + \Omega_r$) close to the natural frequency yielding a combination resonance [25]. For the assumed scaling of the parameters, the steady-state approximated solution consist two dominating harmonic terms: one close to the natural frequency $\Omega_a/2$ and the other at the external force frequency $\Omega_r$.

$$x = a_0 \cos\left(\frac{\Omega_a}{2}\tau - \psi_0\right) + a_r \cos(\Omega_r \tau - \psi_r) + O(\varepsilon^2). \quad (5)$$

The detailed derivation of the parameters $a_0$, $a_r$, $\psi_0$ and $\psi_r$ is provided in [21], and $a_0$ as a function of the various parameters is provided in Appendix A, Eq.(A.1).

The nonlinear stiffness terms have a significant influence on the response and their mutual contribution can be quantified by a single parameter $\kappa_e$, which is the effective nonlinear stiffness:

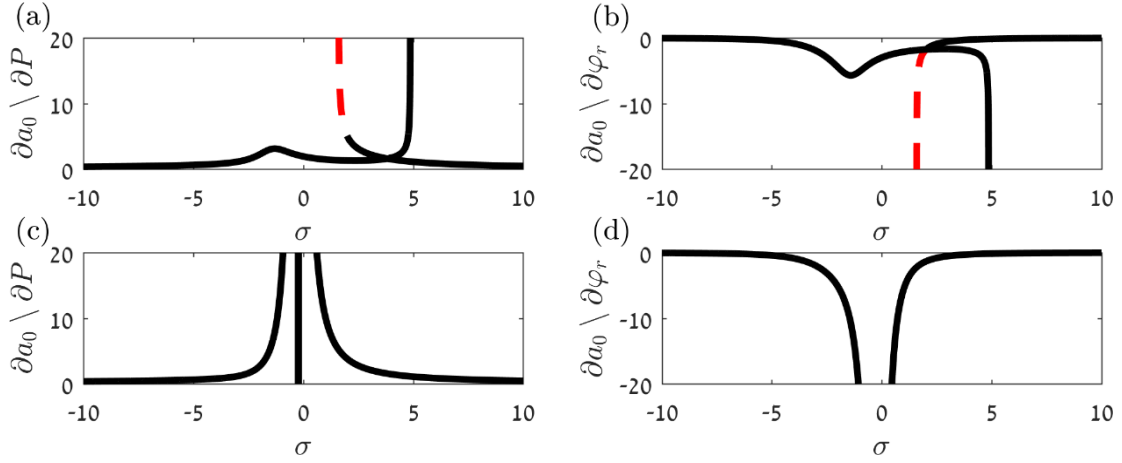

Figure 3: (a) The sensitivity with respect to the external force amplitude, $P$ and (b) with respect to the external phase, $\varphi_r$ when $\kappa_e = 5$. (c) The sensitivity with respect to the external force amplitude, $P$ and (d) with respect to the external phase, $\varphi_r$, when $\kappa_e = 0$. Continuous lines represent stable solutions while dashed lines represent unstable solutions.

$$\kappa_e = 3\kappa_3 - \frac{10}{3}\kappa_2^2. \tag{6}$$

Setting positive effective stiffness leads to a hardening type system, while setting it negative leads to a softening type system [26]. By setting $\kappa_e$ to zero, the amplifier behaves linearly up to the second order (i.e., $\varepsilon^2$). Analytically computed characteristic frequency sweeps for two cases are shown in Figure 2. In the first case, depicted in Figure 2 (a) and (b), the effective stiffness is positive, and one can witness a stiffening behavior. In the second case, depicted in Figure 2 (c) and (d), the effective stiffness equals zero, and the amplifier behaves pseudo-linearly.

Another important attribute of the amplifier is its sensitivity to variations in the external force amplitude, and phase $\varphi_r$, which can be computed in a closed form as shown in Appendix B of [21]. The sensitivities for the previous cases are depicted in Figure 3. It is noticeable, that when a single stable solution exists (for $\kappa_e = 5$, $\sigma < -2$ or $4.8 < \sigma$, and for $\kappa_e = 0$ the whole $\sigma$ domain) higher sensitivity levels can be achieved when $\kappa_e = 0$. Observing Figure 2 and Figure 3, it is also noticeable that the pseudo-linear amplifier sensitivities and amplitude $a_0$ behave similarly with respect to $\sigma$, which means that no compromise has to be made between the two, unlike when $\kappa_e \neq 0$.

## 2.1 Optimal parameters tuning

By properly setting the tunable parameters of the scheme parameters, the oscillator becomes as a sensitive amplifier. To produce large amplification while maintaining the sensitivities, it was found that the effective stiffness should be tuned to zero, therefore, it is suggested to set $\kappa_2 = 1$ and $\kappa_3 = 10/9$. The pumping magnitudes should be tuned to allow the linear system to lose its stability, thus producing large amplitudes. The suitable value of $\gamma_a$ as a function of the system parameters and $\gamma_b$ can be computed by solving a sixth order polynomial:

$$G_6(\varepsilon, \Omega_r)\gamma_a^6 + G_4(\varepsilon, \Omega_r, \gamma_b, \zeta)\gamma_a^4 + G_2(\varepsilon, \Omega_r, \gamma_b, \zeta)\gamma_a^2 + G_0(\Omega_r, \zeta) = 0. \tag{7}$$

The functions $G_i$ are provided in Appendix A of [21]. Because $\gamma_b$ is unknown prior to the computation of $\gamma_a$, one can set the maximum allowable stiffness modulation ($\Delta k_1$, $0 < \Delta k_1 < 1$) as a design parameter. Then, the following algebraic relation can be substituted to Eq.(7):

$$\varepsilon(\gamma_a + \gamma_b) = \Delta k_1, \tag{8}$$

for simple computation of the pumping magnitudes.

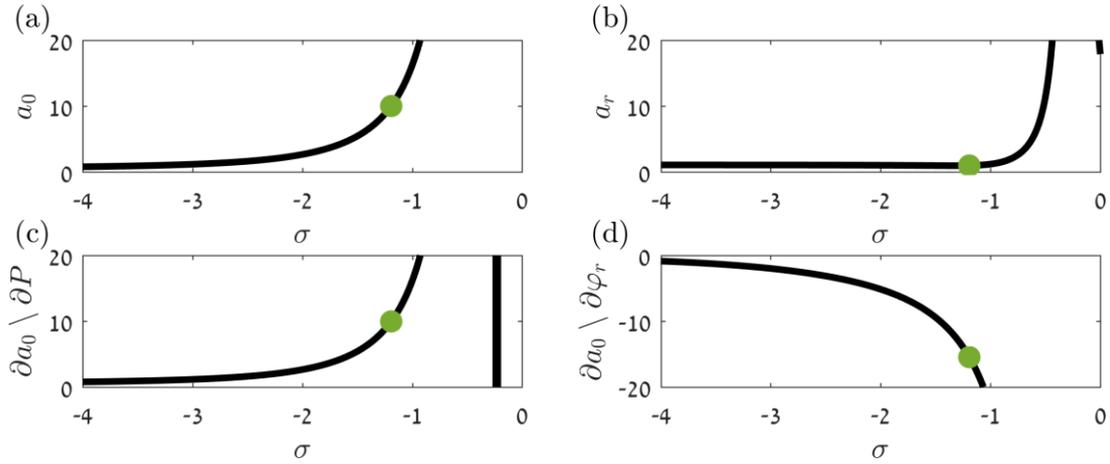

Figure 4: $\kappa_e =0$, the response amplitudes and sensitivities vs. the detuning parameter. (a) The amplitude of the harmonic term near the natural frequency and (b) at the external force frequency. (c) The sensitivity with respect to the external force amplitude, and (d) with respect to the external phase. The dots mark $\sigma \approx -1.19$.

Once the nonlinear stiffness terms and pumping magnitudes were computed the detuning parameter $\sigma$ should be set. This can be done graphically, as shown in Figure 4, where the selected detuning parameter was chosen to be −1.19, and is marked in the subfigures by a circular marker. Finer adjustments can then be made by tuning $\varphi_a$ [21], while $\varphi_b$ is used to obtain $\varphi_r$ as briefly explained in Section 3; a detailed explanation is provided in [16].

## 3   Suggested mass balancing procedure exploiting the amplifier

During the MBP one seeks the distribution of the imbalance forces, amplitude and phase along the rotor. To simplify the discussion, consider the case of an unbalanced rigid disc mounted on a flexible weightless shaft as shown in Figure 5. As depicted in the figure, the center of mass does not coincide with the geometrical center, and is located at a phase $\varphi_r$ relative to the initial position.

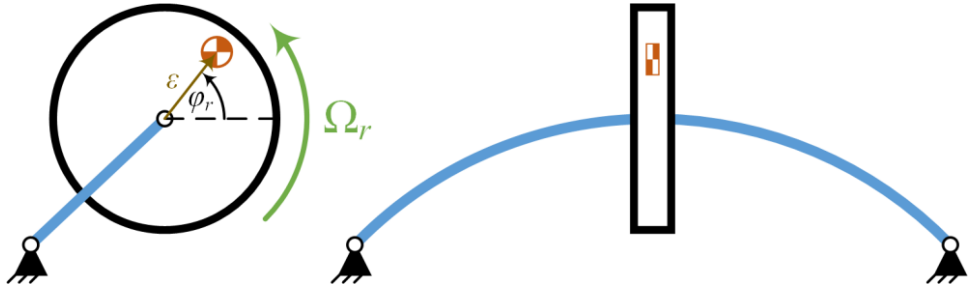

Figure 5: Two views of an unbalanced rigid disc mounted on a weightless flexible shaft.

The resulting imbalance force has the general form $A\Omega_r^2 \cos(\Omega_r t + \varphi_r)$ where $\Omega_r$ is the rotation speed and $A$ is related to the imbalance magnitude. When the SDOF scheme is implemented (i.e., tuned parametric excitation), and the amplitude of vibration close to the natural frequency due to the imbalance is plotted as a function of $\varphi_b + \varphi_r$, it has a period of $\pi$, as shown analytically and numerically in Figure 6.

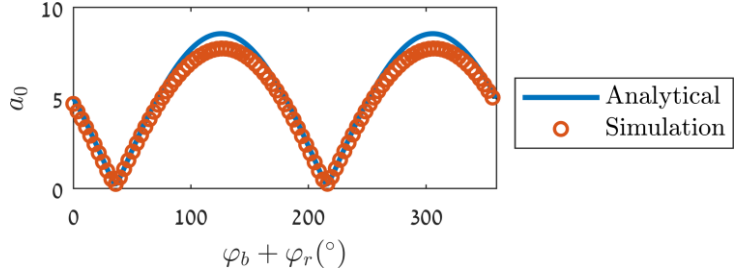

Figure 6: The amplitude of the harmonic term near the natural frequency vs. the sum of phases $\varphi_b+\varphi_r$. Analytically computed (continuous lines) and numerically simulated responses (hollow circular markers).

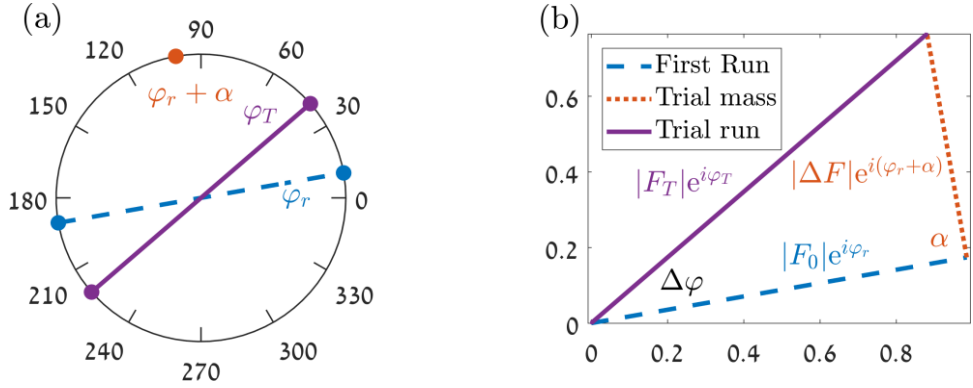

Figure 7: (a) Possible solutions for $\varphi_r$ in the first and the trial runs. (b) Extraction of the imbalance.

Because $\varphi_b$ and $\varphi_r$ have the same influence on the system [16,18,19,21], and $\varphi_b$ is controllable, identifying the minima (or maxima) of $a_0$ vs. $\varphi_b$ is equivalent to identifying the imbalance phase. However, because the two minima are $\pi$ radians apart, an additional, controlled measurement is required, where a trial mass is placed at a known phase.

First, the amplitude vs. the phase $\varphi_b$ is measured until a minimum is found, this phase is denoted $\varphi_{b0}$. The location of the imbalance, $\varphi_r$ is found according to the analytical solution of $a_0(\varphi_{b0}+\varphi_r)$, (e.g., for the SDOF discussed in Section 2 see Eq.(A.1), and for a MDOF system see Eq.(16) and Eq.(20)). To distinguish between the two possible locations of $\varphi_r$ ($\pi$ radians apart, as shown in Figure 7(a)), a trial mass is added, and the total imbalance force at the trial run is:

$$|F_T|e^{i\varphi_T} = |F_0|e^{i\varphi_r} + |\Delta F|e^{i(\varphi_r+\alpha)}, \qquad (9)$$

whereas $|F_0|$ is the force due to the unknown imbalance, $|\Delta F|$ is the force due to the trial mass, and $\varphi_r+\alpha$ is the total trial mass phase. The phase $\varphi_T$ is found by sweeping $\varphi_b$ (two possible solutions $\pi$ radians apart, as shown in Figure 7(a)). The trial mass should be placed about $\pi/2$ apart from $\varphi_r$ to achieve maximum change in $\varphi_r$. Since the imbalance at the trial run must lie between $\varphi_r$ and $\varphi_r+\alpha$, the true location of $\varphi_r$ can be found. And the imbalance magnitude is computed as:

$$|F_0| = \frac{|\Delta F|\sin(\alpha+\Delta\varphi)}{\sin(\Delta\varphi)}, \qquad (10)$$

graphical explanation is depicted in Figure 7(b).

## 4 Extension of the method to a rotating MDOF system

In this section, the scheme is extended to accommodate rotating structures via a test case, for which an experimental rig was designed and built. As shown in Figure 8, two rigid discs are fixed to a rigid rotor which is mounted on a plate free to move only in the horizontal plane (x-z). A lumped parameters model of

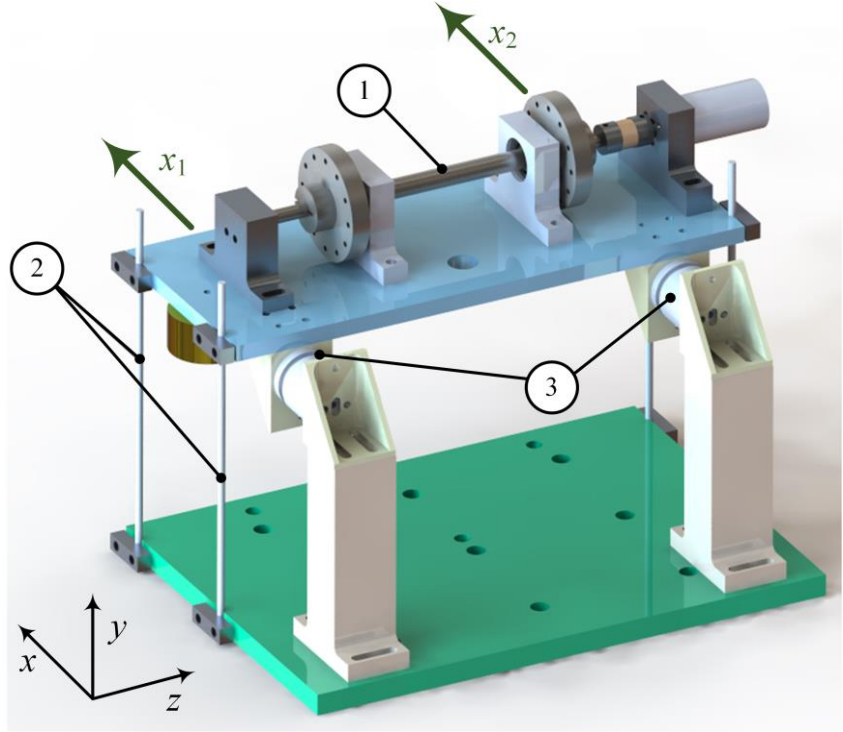

Figure 8: Rotor balancing demonstrator, comprises a rotating rigid shaft and discs (1), flexible foundation supports (2), and two linear voice coil actuators (3) that induces position and time dependent forces.

the system is used, and the governing equations of motion in matrix and vector notation are:

$$\mathbf{M}\ddot{\mathbf{x}} + \mathbf{C}\dot{\mathbf{x}} + \left(\mathbf{K} + \mathbf{K_p}(t)\right)\mathbf{x} + \mathbf{f_{nl}}\left(\mathbf{x}^2, \mathbf{x}^3\right) = \Omega_r^2 \mathbf{f_{ib}} \tag{11}$$

Here, $\mathbf{M}$ is the mass matrix, $\mathbf{C}$ the damping matrix, $\mathbf{K}$ the stiffness matrix, $\mathbf{K_p}$ the pumping matrix, $\mathbf{f_{nl}}$ the nonlinear feedback force vector and $\Omega_r^2 \mathbf{f_{ib}}$ is the imbalance force vector. In this case, the matrix $\mathbf{K_p}$ and force vector $\mathbf{f_{nl}}$ are fully controllable and are manipulated to excite every desired system vibration mode as described below.

By assuming low damping levels, setting the parametric excitation and nonlinear forces to the same order of magnitude, and using the mass normalized modes of the linear undamped system, the equations of motion can be transformed to the following:

$$\mathbf{I}\boldsymbol{\eta}'' + \boldsymbol{\chi}^2 \boldsymbol{\eta} = \mathbf{P}(\tau) - \varepsilon\left(2\zeta\chi\boldsymbol{\eta}' + \mathbf{K_\gamma}(\tau)\boldsymbol{\eta} + \boldsymbol{\kappa_2}\tilde{\boldsymbol{\eta}}_2\right) - \varepsilon^2 \boldsymbol{\kappa_3}\tilde{\boldsymbol{\eta}}_3. \tag{12}$$

Here the time was scaled as $\tau = \tilde{\omega}t$, and $\boldsymbol{\eta}$ is the modal coordinate vector which is related to the physical ones according to Eq.(B.4). A detailed mathematical description regarding the transformation from Eq.(11) to Eq.(12), and parameter definitions are provided in Appendix B.

## 4.1 Approximate analytical solution

To derive the approximate analytical solution, the method of multiple scales is employed. First the imbalance force projection on the first mode is computed, hence $\tilde{\omega} = \omega_1$. It proves convenient to apply the controlled forces in a manner leading to minimal coupling between the modes, while choosing the frequencies leading to combination and principal parametric resonances. Therefore, the following parametric excitation matrix is used:

$$\mathbf{K}_\gamma(\tau) = \begin{pmatrix} \chi_1^2\left(\gamma_{a1}\cos(\chi_{a1}\tau+\varphi_{a1})+\gamma_{b1}\cos(\chi_{b1}\tau+\varphi_{b1})\right) & 0 \\ 0 & 0 \end{pmatrix}, \quad (13)$$

$$\chi_{a1} = 2(1+\varepsilon\sigma), \quad \chi_{b1} + \mu_1\chi_1 = 1+\varepsilon\sigma, \quad \chi_r = \mu_1\chi_1.$$

Moreover, the nonlinear forces should also have minimal influence on the unexcited mode:

$$\mathbf{\kappa}_2 = \frac{|\Phi|^2 \beta_{11}}{\tilde{\omega}^2}\begin{pmatrix} \phi_{11} & 0 & -\phi_{12} \\ \phi_{22}^2 & & \phi_{21}\phi_{22} \\ 0 & 0 & 0 \end{pmatrix}, \quad \mathbf{\kappa}_3 = \frac{|\Phi|^2 \delta_{11}}{\tilde{\omega}^2}\begin{pmatrix} \phi_{11}^2 & 0 & -\frac{3\phi_{11}\phi_{12}}{\phi_{21}\phi_{22}} & -\frac{\phi_{12}(\phi_{12}\phi_{21}+\phi_{11}\phi_{22})}{\phi_{21}^2\phi_{22}} \\ \phi_{22}^2 & & & \\ 0 & 0 & 0 & 0 \end{pmatrix}, \quad (14)$$

$$|\Phi|^2 = (\phi_{12}\phi_{21} - \phi_{11}\phi_{22})^2.$$

Following the multiple scales procedure up to the 2$^{nd}$ order, as detailed in [21], the solution in modal coordinates is:

$$\eta_1 = a_{10}\cos\left(\frac{\chi_{a1}}{2}\tau - \psi_0\right) + a_{r1}\cos(\chi_r\tau - \psi_{r1}), \quad \eta_2 = a_{r2}\cos(\chi_r\tau - \psi_{r2}). \quad (15)$$

Whereas,

$$a_{10} = \frac{\gamma_{b1}P_1}{2\chi_1^2(\mu_1^2-1)}\begin{pmatrix} 4\varepsilon\zeta_1\left(3+\mu_1\left(\mu_1+\mu_1^2-13\right)\right)\cos(\psi_0+\varphi_{b1}+\varphi_1) + \\ (1+\mu_1)\begin{pmatrix} \gamma_{a1}\varepsilon\left(7+(\mu_1-4)\mu_1\right)\sin(\psi_0+\varphi_{a1}-\varphi_{b1}-\varphi_1) + \\ 4(\mu_1-3)(\mu_1-1)(2-\varepsilon\sigma)\sin(\psi_0+\varphi_{b1}+\varphi_1) \end{pmatrix} \end{pmatrix} \quad (16)$$
$$\times\left(4(\mu_1-3)(\mu_1-1)(\mu_1+1)(4\zeta_1+\gamma_{a1}(1-\varepsilon\sigma)\sin(2\psi_0+\varphi_{a1}))\right)^{-1}$$

It is clear that Eq.(16) is practically the same as Eq.(A.1), therefore the optimal parameters can be tuned as described in Section 2.1. Furthermore, the suggested balancing procedure (Section 3) can be employed and the imbalance projection on the first mode can be found.

To find the imbalance projection on the second mode, the parameters are tuned to have minimal influence on the first mode. In this case, $\tilde{\omega} = \omega_2$ and the pumping and nonlinear terms are tuned as:

$$\mathbf{K}_\gamma(\tau) = \begin{pmatrix} 0 & 0 \\ 0 & \chi_2^2\left(\gamma_{a2}\cos(\chi_{a2}\tau+\varphi_{a2})+\gamma_{b2}\cos(\chi_{b2}\tau+\varphi_{b2})\right) \end{pmatrix}, \quad (17)$$

$$\chi_{a2} = 2(1+\varepsilon\sigma), \quad \chi_{b2} + \mu_2\chi_1 = 1+\varepsilon\sigma, \quad \chi_r = \mu_2\chi_1.$$

$$\mathbf{\kappa}_2 = \frac{|\Phi|^2 \beta_{22}}{\tilde{\omega}^2}\begin{pmatrix} 0 & 0 & 0 \\ -\frac{\phi_{21}}{\phi_{11}\phi_{12}} & 0 & \frac{\phi_{22}}{\phi_{11}^2} \end{pmatrix}, \quad \mathbf{\kappa}_3 = \frac{|\Phi|^2 \delta_{22}}{\tilde{\omega}^2}\begin{pmatrix} 0 & 0 & 0 & 0 \\ -\frac{\phi_{21}(\phi_{12}\phi_{21}+\phi_{11}\phi_{22})}{\phi_{11}\phi_{12}^2} & -\frac{3\phi_{21}\phi_{22}}{\phi_{11}\phi_{12}} & 0 & \frac{\phi_{22}^2}{\phi_{11}^2} \end{pmatrix}\delta_{22}. \quad (18)$$

And the solution is:

$$\eta_1 = a_{r2}\cos(\chi_r\tau - \psi_{r2}), \quad \eta_2 = a_{20}\cos\left(\frac{\chi_{a2}}{2}\tau - \psi_0\right) + a_{r2}\cos(\chi_r\tau - \psi_{r2}). \quad (19)$$

Whereas,

$$a_{20} = \frac{\gamma_{b2}P_2}{2\chi_1^2(\mu_2^2-1)}\begin{pmatrix} 4\varepsilon\zeta_2\left(3+\mu_2\left(\mu_2+\mu_2^2-13\right)\right)\cos(\psi_0+\varphi_{b2}+\varphi_2) + \\ (1+\mu_2)\begin{pmatrix} \gamma_{a2}\varepsilon\left(7+(\mu_2-4)\mu_2\right)\sin(\psi_0+\varphi_{a2}-\varphi_{b2}-\varphi_2) + \\ 4(\mu_2-3)(\mu_2-1)(2-\varepsilon\sigma)\sin(\psi_0+\varphi_{b2}+\varphi_2) \end{pmatrix} \end{pmatrix} \quad (20)$$
$$\times\left(4(\mu_2-3)(\mu_2-1)(\mu_2+1)(4\zeta_2+\gamma_{a2}(1-\varepsilon\sigma)\sin(2\psi_0+\varphi_{a2}))\right)^{-1}.$$

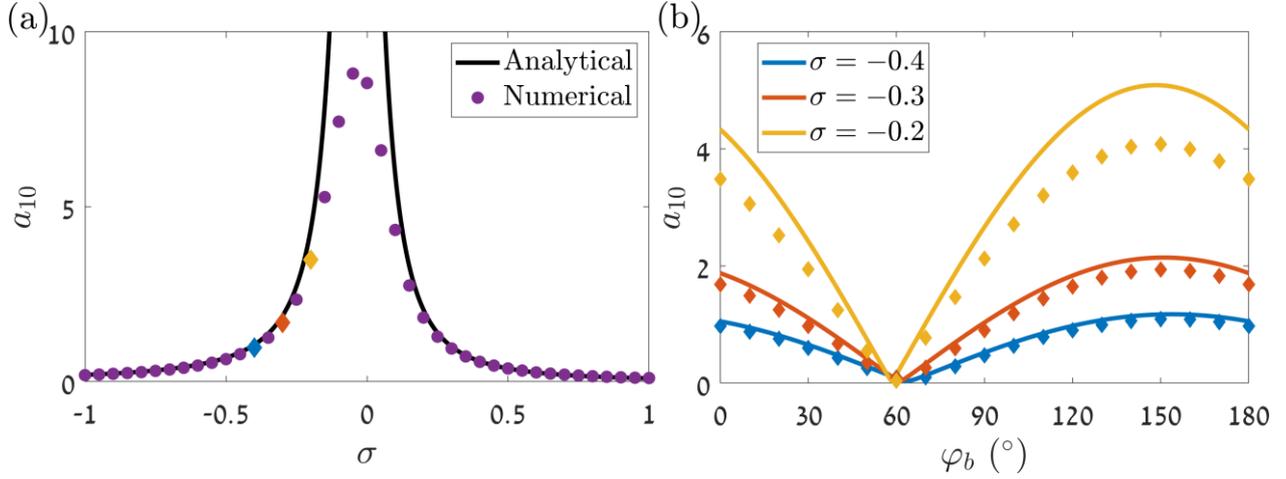

Figure 9: first mode amplitude close to the natural frequency ($a_{10}$) (a) vs. $\sigma$, and (b) vs. $\varphi_b$. Analytical solutions are shown by continous lines, and numerical results by markers.

As before, Eq.(20) is practically the same as Eq.(A.1). The optimal parameters can be tuned as previously described, and the suggested balancing procedure can be employed.

### 4.2 Numerical verification

In the following section a numerical verfication of the suggested optimal schem is performed via simulations. The parameters used in the simulation match the experimental rig shown in Figure 8, and were experimentally identified:

$$\Phi = \begin{pmatrix} 0.6411 & 0.6312 \\ 0.6230 & -0.6645 \end{pmatrix}, \quad \Omega_n = \begin{pmatrix} 18.9 & 0 \\ 0 & 29.1 \end{pmatrix} (\text{Hz}), \quad \zeta = \begin{pmatrix} 1 & 0 \\ 0 & 0.45 \end{pmatrix}. \tag{21}$$

The numerical verification shown herein was performed for a known level of imbalance. The goal was to find its projection on each mode, and to study the ability to change the sensitivity by tuning the pumping frequency, $\sigma$.

First, the imbalance projection on the first mode was sought, therefore the parameters were tuned according to Eqns.(13) and (14), where $\kappa_{2,11}$ and $\kappa_{3,11}$ were chosen according to Eq.(6) leading to $\kappa_e=0$. Moreover, the pumping magnitudes $\gamma_{a1}$ and $\gamma_{b1}$ were tuned according to Eqns.(7) and (8), where $\Delta k=0.1$. In Figure 9 (a) the analytical and numerical results of the amplitude close to the natural frequency ($a_{10}$) vs. the detuning parameter $\sigma$ are shown. The analytical solution is shown by a black continuous line, while, the numerical results are depicted by markers. The values depicted by diamond shaped markers in Figure 9 (a) are used in the following stage, the phase ($\varphi_b$) sweep. One can witness the good agreement between the analytical and numerical solutions for relatively low amplitudes. As the amplitude grows the solutions deviate due to nonlinear effects, this is anticipated whereas the analytical model was derived for small amplitudes. Moreover, it is noticeable that according to the analytical model a single solution exists at each frequency, in contrast to the previous method. A discussion regarding that is provided in Section 4.3.

The following stage is a frequency sweep as shown in Figure 9 (b) for different detuning values. The numerical solutions (markers) are in agreement with the analytical solution (continuous lines), and the agreement is better as the amplitude decreases. Because the numerical solution agrees with the model, the imbalance phase $\varphi_r$ can be computed from Eq.(16). As discussed in Section 2, the amplification and sensitivity behave similarly with respect to $\sigma$ when the equivalent stiffness is set to zero. This phenomenon can be seen in Figure 9 (b); as the detuning parameter is increased higher amplitudes are produced and also better sensitivity is achieved. The sensitivity can be interpreted as the difference between the maximum and minimum measured amplitudes in the phase sweep.

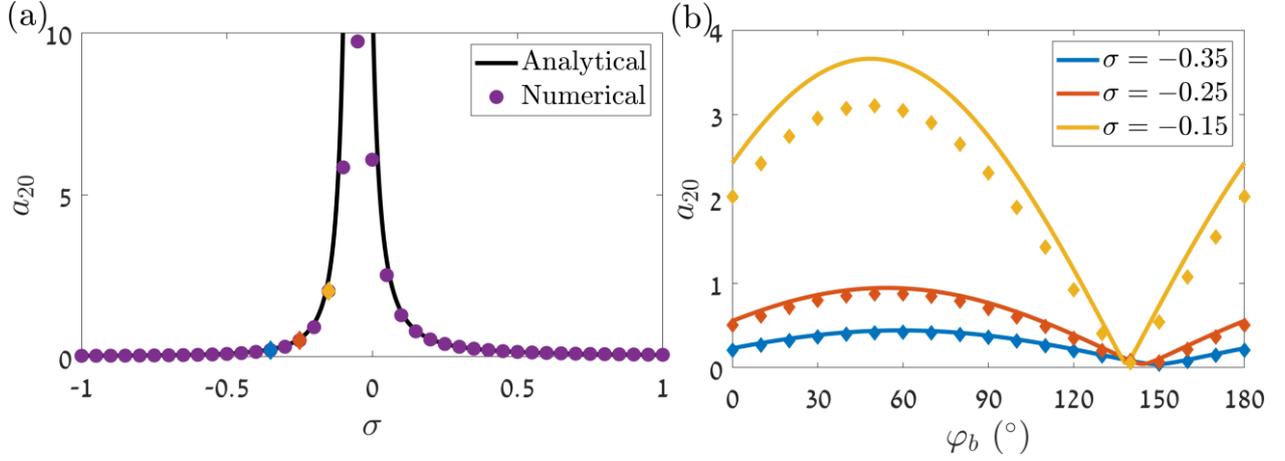

Figure 10: second mode amplitude close to the natural frequency ($a_{10}$) (a) vs. $\sigma$, and (b) vs. $\varphi_b$. Analytical sulutions are shown by lines, and numerical results by markers.

A similar process to identify the imbalance projection on the second mode was preformed, and shown in Figure 10. For this case, the parameters were tuned according to Eqns.(17) and (18) where $\kappa_{2,21}$ and $\kappa_{3,21}$ were chosen according to Eq.(6) leading to $\kappa_e=0$. Moreover, the pumping magnitudes $\gamma_{a1}$ and $\gamma_{b1}$ were tuned according to Eqns.(7) and (8), where $\Delta k=0.1$. The observations in this case are similar to the ones in the previous case, hence the imbalance phase $\varphi_r$ can be computed from Eq.(20), and the amplification and sensitivity are controllable via $\sigma$.

### 4.3   Comparison to the previous MBP method

The previous MBP method, which is discussed in [15-20], differs from the current by two main aspects. It neither includes a quadratic stiffness term, nor it includes the parameter optimization stage for selecting the parametric excitation magnitudes. Therefore, the previous method has two main drawbacks: (1) a compromise between amplification and sensitivity has to be made, and (2) multiple stable solution branches may exist very close to each other for the same $\sigma$. In practice, two close branches results in a major decrease of the sensitivity, since the solution jumps from one stable branch to the other as shown Figure 11 (and [16]), by the circular and diamond shaped markers. These results, depicting two close stable solution branches were obtained experimentally.

By implementing the quadratic nonlinear stiffness, the system behavior is pseudo-linear, hence omits the need to compromise between sensitivity and amplification, and it also eliminates the second stable solution. The use of optimized pumping magnitudes enables higher sensitivity, and also results with a single stable solution, because the pumping level of the principal parametric resonance is lower than its linear threshold [21].

Comparison between the two methods applied to the same system [16] is shown in Figure 11. It is noticeable that while similar maximal amplitudes were produced, the sensitivity is about five times higher when the optimal method is used. Note that the methods result in different values for $\varphi_{b0}$, since the phase of the response $\psi$ differs. For the experimental results, $\varphi_{b0}$ is located at the minimum of the black dashed line (~84°), while for the optimal method it is located around 60°.

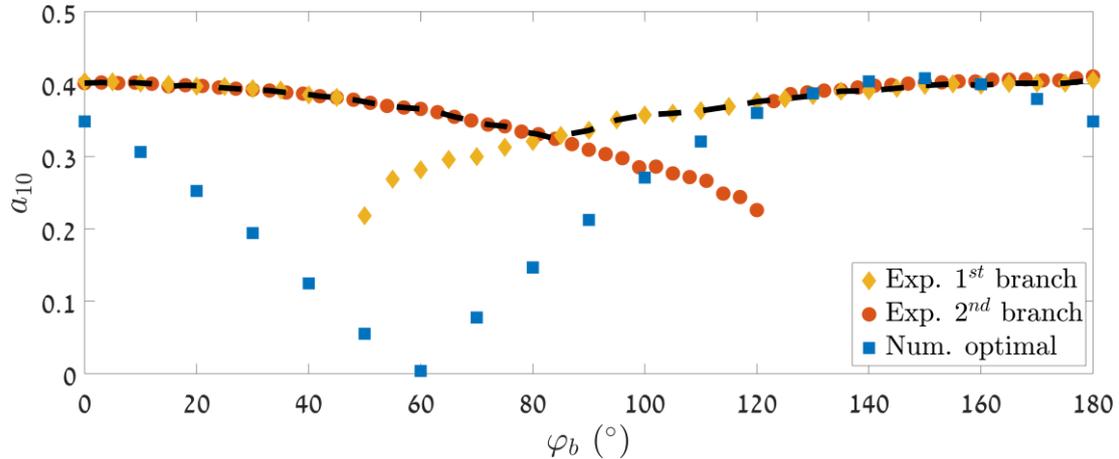

Figure 11: first mode amplitude close to the natural frequency ($a_{10}$) vs. $\varphi_b$, a comparison between the current optimal to the previous metod. Previous method, experimentally obtained results: circular orange (diamond shaped yellow) markers depict the first (second) stable solution branch. Current method, numerically obtained results are shown by blue square markers.

## 5  Summary and conclusions

A modified optimal mass balancing procedure scheme has been introduced, whose performance surpassed the previous one [16], as discussed in Section 4.3. The current scheme is an extension of the SDOF parametric amplifier scheme, which was discussed in Section 2, to a rotating MDOF system. The current MBP utilizes tuned dual frequency parametric excitation and nonlinear feedback term, to project the distributed imbalance force on any selected vibration mode, while rotating the structure slowly. The latter enables to identify the imbalance and correct it by adding or removing mass from the rotor, while avoiding the difficulties and risks involved in rotating the rotor close to a critical speed.

Implementing tuned dual parametric excitation leads to two kind of resonances. First, a combination resonance that couples the external signal (i.e., the imbalance force) with one of the vibration modes, which is appropriate to a selected natural frequency. Second, a principal parametric resonance that considerably amplifies the response close to the selected natural frequency. The resulting dynamic response due to both resonances is the underlying mechanism that allows energy transfer from the imbalance force at frequency $\omega_r$ to the rotor's response close to a selected natural frequency ~ $\omega_n$. The optimal parameters tuning by which the pumping magnitudes are selected, leads to a marginally stable system [21], thus making it very sensitive. In addition, proper tuning of the nonlinear feedback term (i.e., setting the equivalent stiffness to zero) leads to a pseudo-linear dynamic behavior. As a result, the amplification and sensitivity behaves similarly with respect to the detuning parameter $\sigma$, therefore no compromise between the two has to be made as in the previous method.

In Section 4.3, a comparison between the optimal and previous methods was briefly discussed. The data for the previous method was obtained experimentally, while for the current method it was simulated. In future work, the optimal method would be implemented experimentally on the same system to validate its superiority. In addition, the method will be tested on a different system which has "flexible" modes [27].

## Acknowledgements

The first author would like to acknowledge the generous financial support of the Israeli Ministry of Science and Technology (3-13336) for the Applied Science scholarship for engineering PhD students 2016.

# Appendix A

The analytical solution of $a_0$, as reported in [21]:

$$a_0 = \frac{\gamma_b}{2(\Omega_r^2 - 1)} \begin{pmatrix} 4\varepsilon\zeta\left(3 + \Omega_r\left(\Omega_r^2 + \Omega_r - 13\right)\right)\cos(\psi_0 + \varphi_b + \varphi_r) + \\ (1+\Omega_r)\begin{pmatrix} \gamma_a\varepsilon\left(7 + (\Omega_r - 4)\Omega_r\right)\sin(\psi_0 + \varphi_a - \varphi_b - \varphi_r) + \\ 4(\Omega_r - 3)(\Omega_r - 1)(2 - \varepsilon\sigma 2)\sin(\psi_0 + \varphi_b + \varphi_r) \end{pmatrix} \end{pmatrix} \\ \times \left(4(\Omega_r - 3)(\Omega_r - 1)(\Omega_r + 1)\left(4\zeta + \gamma_a(1-\varepsilon\sigma)\sin(2\psi_0 + \varphi_a)\right)\right)^{-1}.$$

(A.1)

Whereas, $\psi_0$ are the roots of a 10[th] order polynomial, whose coefficient are omitted for brevity.

# Appendix B

The governing equations of the experimental rig are:

$$\underbrace{\begin{pmatrix} m_{11} & m_{12} \\ m_{21} & m_{22} \end{pmatrix}}_{\mathbf{M}}\begin{pmatrix} \ddot{x}_1 \\ \ddot{x}_2 \end{pmatrix} + \underbrace{\begin{pmatrix} c_{11} & c_{12} \\ c_{21} & c_{22} \end{pmatrix}}_{\mathbf{C}}\begin{pmatrix} \dot{x}_1 \\ \dot{x}_2 \end{pmatrix} + \underbrace{\begin{pmatrix} k_{11} & k_{12} \\ k_{21} & k_{22} \end{pmatrix}}_{\mathbf{K}}\underbrace{\begin{pmatrix} x_1 \\ x_2 \end{pmatrix}}_{\mathbf{x}} + \\ \underbrace{\begin{pmatrix} k_{p11}(t) & k_{p12}(t) \\ k_{p21}(t) & k_{p22}(t) \end{pmatrix}}_{\mathbf{K_p}}\begin{pmatrix} x_1 \\ x_2 \end{pmatrix} + \underbrace{\begin{pmatrix} b_{11} & b_{12} \\ b_{21} & b_{22} \end{pmatrix}}_{\mathbf{B}}\begin{pmatrix} x_1^2 \\ x_2^2 \end{pmatrix} + \underbrace{\begin{pmatrix} d_{11} & d_{12} \\ d_{21} & d_{22} \end{pmatrix}}_{\mathbf{D}}\begin{pmatrix} x_1^3 \\ x_2^3 \end{pmatrix} = \Omega_r^2 \underbrace{\begin{pmatrix} f_1 \cos(\Omega_r t + \varphi_I) \\ f_2 \cos(\Omega_r t + \varphi_{II}) \end{pmatrix}}_{\mathbf{Q}}$$

(B.1)

It is assumed that using the normal modes of the linear system, the parametric excitation matrix can be made diagonal:

$$\mathbf{\Phi}^T \mathbf{K_p} \mathbf{\Phi} = \mathbf{K_{p\eta}},$$

(B.2)

whereas, $\mathbf{\Phi}$ is the modal matrix of the linear undamped system, containing the normal modes of the system in the following form:

$$\mathbf{\Phi} = \begin{pmatrix} \phi_1 & \phi_2 \end{pmatrix}$$

(B.3)

Next, the following transformation is defined:

$$\mathbf{x} = \varepsilon^\alpha \mathbf{\Phi}\mathbf{\eta}, \quad \varepsilon \sim O(\hat{\zeta}) \tag{B.4}$$

Here, $\varepsilon$ is some measure of the modal amplitudes and is assumed small, and $\hat{\zeta}$ is the modal damping. Substituting Eq.(B.4) to Eq.(B.1) and pre-multiplying it by $\mathbf{\Phi}$, the equations reduce to:

$$\mathbf{I}\ddot{\mathbf{\eta}} + 2\hat{\zeta}\mathbf{\Omega}_n\dot{\mathbf{\eta}} + \mathbf{\Omega}_n^2\mathbf{\eta} + \mathbf{K}_{p\eta}(t)\mathbf{\eta} + \varepsilon^\alpha \mathbf{\Phi}^T \mathbf{B}(\mathbf{\Phi}\mathbf{\eta})^2 + \varepsilon^{2\alpha}\mathbf{\Phi}^T\mathbf{D}(\mathbf{\Phi}\mathbf{\eta})^3 = \varepsilon^{-\alpha}\Omega_r^2 \mathbf{\Phi}^T\mathbf{Q}$$

$$\hat{\zeta} = \begin{pmatrix} \hat{\zeta}_1 & 0 \\ 0 & \hat{\zeta}_2 \end{pmatrix} \quad \mathbf{\Omega}_n = \begin{pmatrix} \omega_1 & 0 \\ 0 & \omega_2 \end{pmatrix} \tag{B.5}$$

$$\mathbf{K}_{p\eta}(t) = \begin{pmatrix} \omega_1^2 \begin{pmatrix} \alpha_{a1}\cos(\omega_{a1}t+\varphi_{a1}) + \\ \alpha_{b1}\cos(\omega_{b1}t+\varphi_{b1}) \end{pmatrix} & 0 \\ 0 & \omega_2^2 \begin{pmatrix} \alpha_{a2}\cos(\omega_{a2}t+\varphi_{a2}) + \\ \alpha_{b2}\cos(\omega_{b2}t+\varphi_{b2}) \end{pmatrix} \end{pmatrix}$$

Introducing the dimensionless time $\tau = \tilde{\omega}t$, where $\tilde{\omega}$ is the response typical frequency, to Eq.(B.5):

$$\mathbf{I}\mathbf{\eta}'' + 2\hat{\zeta}\chi\mathbf{\eta}' + \chi^2\mathbf{\eta} + \frac{1}{\tilde{\omega}^2}\mathbf{K}_{p\eta}(\tau)\mathbf{\eta} + \frac{1}{\tilde{\omega}^2}\varepsilon^\alpha \mathbf{\Phi}^T\mathbf{B}(\mathbf{\Phi}\mathbf{\eta})^2 + \frac{1}{\tilde{\omega}^2}\varepsilon^{2\alpha}\mathbf{\Phi}^T\mathbf{D}(\mathbf{\Phi}\mathbf{\eta})^3 = \frac{\Omega_r^2}{\tilde{\omega}^2}\varepsilon^{-\alpha}\mathbf{\Phi}^T\mathbf{Q}$$

$$\chi = \begin{pmatrix} \chi_1 & 0 \\ 0 & \chi_2 \end{pmatrix}, \quad \chi_\bullet = \frac{\omega_\bullet}{\tilde{\omega}} \tag{B.6}$$

Where $\partial/\partial\tau \equiv \bullet'$. Moreover, Eq.(B.6) can be written as Eq.(12) where light damping, weak pumping and large nonlinearity are assumed:

$$\hat{\zeta} = \varepsilon\zeta, \quad \alpha_\bullet = \varepsilon\gamma_\bullet \quad \rightarrow \quad \frac{1}{\tilde{\omega}^2}\mathbf{K}_{p\eta}(\tau) = \varepsilon\mathbf{K}_\gamma(\tau)$$

$$\frac{1}{\tilde{\omega}^2}\varepsilon^\alpha\mathbf{\Phi}^T\mathbf{B}(\mathbf{\Phi}\mathbf{\eta})^2 = \varepsilon\mathbf{\kappa}_2\tilde{\mathbf{\eta}}_2, \quad \tilde{\mathbf{\eta}}_2 = \begin{pmatrix} \eta_1^2 & \eta_1\eta_2 & \eta_2^2 \end{pmatrix}^T$$

$$\frac{1}{\tilde{\omega}^2}\varepsilon^{2\alpha}\mathbf{\Phi}^T\mathbf{D}(\mathbf{\Phi}\mathbf{\eta})^3 = \varepsilon^2\mathbf{\kappa}_3\tilde{\mathbf{\eta}}_3, \quad \tilde{\mathbf{\eta}}_3 \begin{pmatrix} \eta_1^3 & \eta_1^2\eta_2 & \eta_1\eta_2^2 & \eta_2^3 \end{pmatrix}^T$$

$$\mathbf{P}(\tau) = \begin{Bmatrix} p_1\cos(\chi_r\tau+\varphi_1) \\ p_2\cos(\chi_r\tau+\varphi_2) \end{Bmatrix} = \frac{\Omega_r^2}{\tilde{\omega}^2}\varepsilon^{-\alpha}\mathbf{\Phi}^T\mathbf{Q} \tag{B.7}$$

$$p_1 = \varepsilon^{-\alpha}\chi_r^2\sqrt{f_1^2\phi_{11}^2 + f_2^2\phi_{21}^2 + 2f_1f_2\phi_{11}\phi_{21}\cos(\varphi_I - \varphi_{II})}$$

$$p_2 = \varepsilon^{-\alpha}\chi_r^2\sqrt{f_1^2\phi_{12}^2 + f_2^2\phi_{22}^2 + 2f_1f_2\phi_{12}\phi_{22}\cos(\varphi_I - \varphi_{II})}$$

$$\varphi_1 = \arctan\left(\frac{f_1\phi_{11}\sin(\varphi_I) + f_2\phi_{21}\sin(\varphi_{II})}{f_1\phi_{11}\cos(\varphi_I) + f_2\phi_{21}\cos(\varphi_{II})}\right)$$

$$\varphi_2 = \arctan\left(\frac{f_1\phi_{12}\sin(\varphi_I) + f_2\phi_{22}\sin(\varphi_{II})}{f_1\phi_{12}\cos(\varphi_I) + f_2\phi_{22}\cos(\varphi_{II})}\right)$$